\documentclass[10pt,twocolumn,letterpaper]{article}

\usepackage[pagenumbers]{iccv}   

\definecolor{iccvblue}{rgb}{0.21,0.49,0.74}
\usepackage[pagebackref,breaklinks,colorlinks,allcolors=iccvblue]{hyperref}
\usepackage{enumitem} 
\usepackage{indentfirst} 
\usepackage{graphicx}
\usepackage{float}
\usepackage{bm}
\usepackage{multirow}
\usepackage{booktabs}
\usepackage{amssymb}
\usepackage{array}
\usepackage{mathtools}
\usepackage{comment}
\usepackage{colortbl} 
\usepackage{xcolor}   

\usepackage[hypcap=false]{caption}

\def\paperID{6046} 
\def\confName{ICCV}
\def\confYear{2025}

\title{IM-LUT: Interpolation Mixing Look-Up Tables for Image Super-Resolution}

\author{Sejin Park\hspace{1cm} Sangmin Lee\hspace{1cm}  Kyong Hwan Jin\hspace{1cm} Seung-Won Jung\thanks{corresponding author}\\
{Department of Electrical Engineering, Korea University}\\
{\tt\small \{sejin5485, etsq1222, kyong\_jin, swjung83\}@korea.ac.kr}
}

\begin{document}
\maketitle
\begin{abstract}

Super-resolution (SR) has been a pivotal task in image processing, aimed at enhancing image resolution across various applications. Recently, look-up table (LUT)-based approaches have attracted interest due to their efficiency and performance. However, these methods are typically designed for fixed scale factors, making them unsuitable for arbitrary-scale image SR (ASISR). Existing ASISR techniques often employ implicit neural representations, which come with considerable computational cost and memory demands. To address these limitations, we propose Interpolation Mixing LUT (IM-LUT), a novel framework that operates ASISR by learning to blend multiple interpolation functions to maximize their representational capacity. Specifically, we introduce IM-Net, a network trained to predict mixing weights for interpolation functions based on local image patterns and the target scale factor. To enhance efficiency of interpolation-based methods, IM-Net is transformed into IM-LUT, where LUTs are employed to replace computationally expensive operations, enabling lightweight and fast inference on CPUs while preserving reconstruction quality. Experimental results on several benchmark datasets demonstrate that IM-LUT consistently achieves a superior balance between image quality and efficiency compared to existing methods, highlighting its potential as a promising solution for resource-constrained applications. 
\end{abstract}
\section{Introduction}
\label{sec:intro}

Image super-resolution (SR), a fundamental task in low-level computer vision, aims to transform low-resolution (LR) images into high-resolution (HR) counterparts. SR plays a crucial role in a wide range of applications, including medical imaging, autonomous driving, and display technologies. Despite recent advances in deep learning-based SR, their high computational complexity and memory demands hinder deployment in resource-limited environments. To overcome these limitations, efficient SR solutions have been proposed, focusing on simplifying model structures~\cite{shi2016real, ahn2018fast}, and adopting techniques like low-bit quantization~\cite{li2020pams, lee2024refqsr}. 
\begin{figure}[!t] 
\centering
\begin{minipage}{1.0\linewidth}
\centerline{\includegraphics[width=0.98\textwidth, height=0.31\textheight]{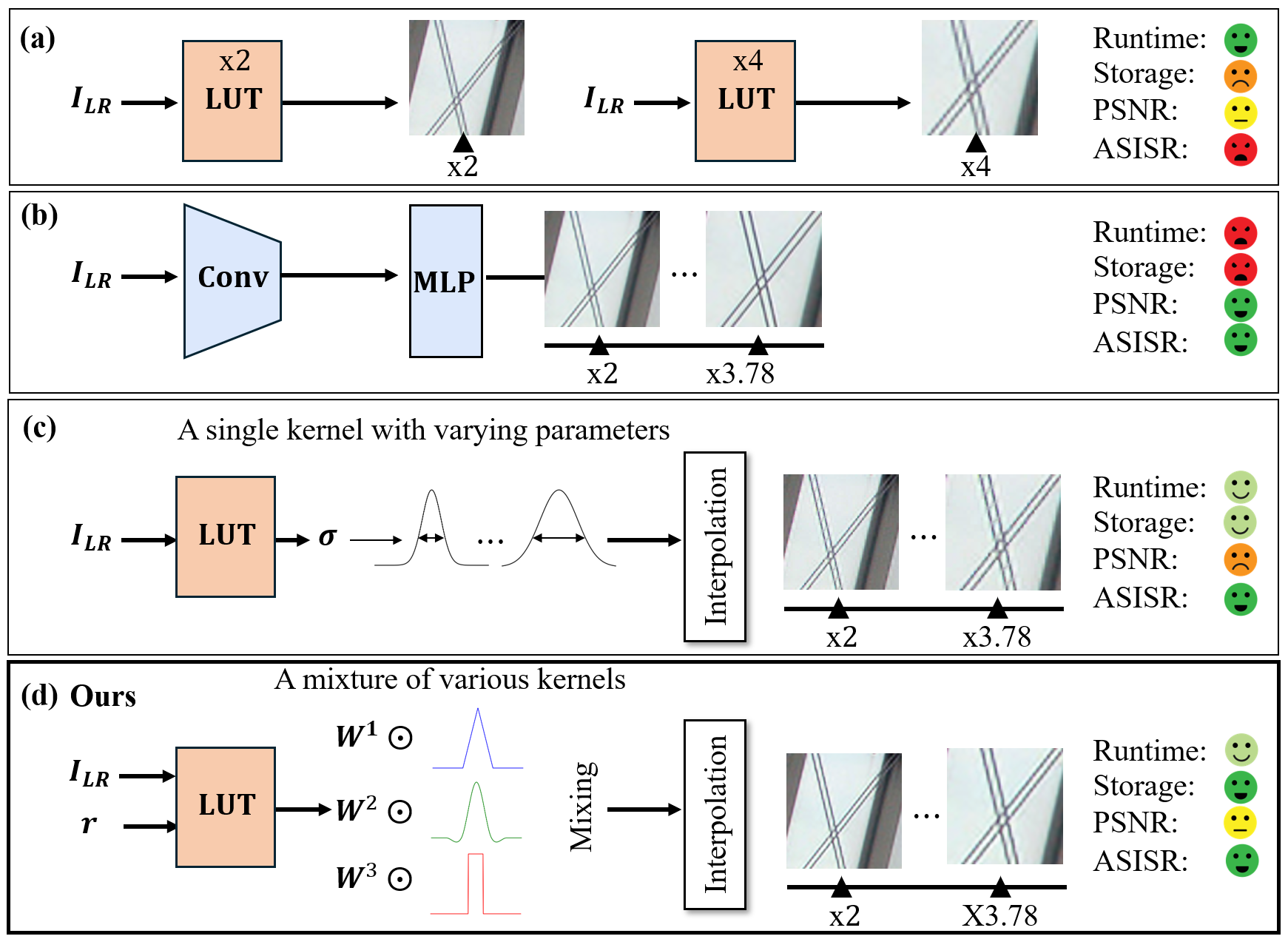}}
\end{minipage}

\caption{Comparison of SR approaches: (a) Conventional LUT-based methods~\cite{jo2021practical, li2022mulut, li2024look} support only fixed-scale SR, (b) the INR-based methods~\cite{chen2021learning, lee2022local, he2024latent} support ASISR but with high computational costs, and both (c) LeRF~\cite{li2023learning} and (d) our proposed IM-LUT support ASISR using LUTs. Unlike LeRF, which employs a single-kernel interpolation with varying parameters, IM-LUT combines various interpolation kernels, achieving higher representation capacity with lower computational costs.}
\label{fig_motivation}
\end{figure}

Meanwhile, look-up table (LUT)-based SR methods have emerged as a promising alternative for achieving extremely efficient SR. By converting a deep SR network to LUTs, these methods enable neural network-free inference. An early LUT approach, SR-LUT~\cite{jo2021practical}, utilized small receptive fields to directly map LR input pixel values to HR output values, significantly reducing computational overhead. Subsequent studies introduced multiple LUT structures and hierarchical mappings to expand the receptive field~\cite{li2022mulut, li2024toward}. 
However, the exponential growth of LUT storage requirements with the expansion of receptive fields imposes inherent limitations on these methods. Moreover, most LUT-based methods remain restricted to fixed-scale SR, hindering their applicability to real-world scenarios that require flexible scaling capabilities. 

Arbitrary-scale image SR (ASISR) enables flexible scaling through scale-adaptive modules. For example, Meta-SR~\cite{hu2019meta} proposed a meta-upscale module that dynamically generates filter weights. LIIF~\cite{chen2021learning} adopted implicit neural representations (INRs) to predict the pixel values in continuous coordinates, enabling precise reconstruction at any desired resolution. However, these methods rely heavily on computationally intensive feature extraction processes, which limits their efficiency. As a result, although they can handle diverse scales, their substantial computational demands and storage requirements hinder their practicality.

To overcome these limitations, we propose interpolation mixing LUT (IM-LUT), a novel framework that performs ASISR by learning to adaptively blend multiple interpolation functions, maximizing their representational capacity. Specifically, we introduce an interpolation mixing network (IM-Net) that predicts pixel-wise weights for multiple interpolation results based on local image content and the target scale factor. By dynamically mixing multiple interpolation results with the transfer from IM-Net to IM-LUT, we achieve robust and compelling SR performance while supporting scale flexibility with minimal computational complexity. The key contributions of this paper are as follows:
\begin{itemize}[label=\textbullet, topsep=0pt, partopsep=0pt, itemsep=0pt, parsep=0pt]
    \item We propose a novel framework called IM-LUT that maximizes the representational capacity of each interpolation function by predicting weights.
    
    \item IM-LUT assigns pixel-wise weights to different interpolation functions based on local texture details and the target scale factor, enabling content-adaptive and scale-adaptive ASISR. 
    \item Extensive experiments on benchmark datasets demonstrate that IM-LUT achieves a competitive trade-off between SR performance and computational efficiency, while enabling fast inference on CPUs and requiring low storage size.
\end{itemize} 

\section{Related Work}
\label{sec:related}

\subsection{Look-Up Table-Based Super Resolution}
SR-LUT~\cite{jo2021practical} pioneered an efficient SR method that transfers a deep network to LUTs, allowing HR output pixel values to be obtained directly by indexing LR input pixel values within a small receptive field. Although SR-LUT enables neural network-free inference but suffers from limited receptive fields due to exponential LUT storage growth. To address this limitation, MuLUT~\cite{li2022mulut, li2024toward} expanded the receptive field using multiple LUTs organized in parallel and hierarchical structures. Furthermore, MuLUT introduced an LUT-aware fine-tuning strategy to refine LUT predictions. SPF-LUT~\cite{li2024look} introduced a novel compression framework, known as diagonal-first compression (DFC), which applies diagonal re-indexing and subsampling strategies to address redundancy in LUTs. However, all the aforementioned LUT-based methods are constrained to fixed-scale SR, limiting their practicality for real-world applications that require flexible zoom-in and zoom-out operations. 

Recently, existing interpolation filters have been integrated to achieve effective LUT-based SR. Notably, LeRF~\cite{li2023learning} used steerable resampling functions, such as Gaussian kernels, combined with a pixel-specific hyper-parameter selection for content-adaptive SR. However, LeRF is constrained by its reliance on a single class of filters, which often fails to capture diverse textural patterns. This limitation highlights the need for more versatile and flexible filtering approaches.

\noindent  \subsection{Arbitrary-Scale Super Resolution}
The field of SR has been advanced in recent years, particularly towards ASISR. ASISR tackles the challenge of upscaling images to arbitrary magnification factors with a single model. Meta-SR~\cite{hu2019meta} introduced the meta-upscale module, which dynamically predicts filter weights for different scales. However, its reliance on scale information solely during the upsampling stage led to uniform feature extraction across scales. To address this, ArbSR~\cite{wang2021learning} incorporated conditional convolutions into the backbone network, enabling features to adapt dynamically to scale-specific characteristics. Moreover, EQSR~\cite{wang2023deep} further integrated scale information directly into both the feature extraction and upsampling processes, effectively addressing challenges posed by scale variation. These methods have demonstrated robust performance across different scale factors. More recently, SLFAM and LFAUM~\cite{zhao2024activating} enhanced SR capabilities by incorporating scale and local feature information through adaptive receptive fields.

Advancements in ASISR have also extended into the domain of INR. As a pioneering approach, LIIF~\cite{chen2021learning} presented a framework that predicts pixel values for any coordinate by leveraging spatially adjacent feature representations and coordinate-based queries. Furthermore, LTE~\cite{lee2022local} proposed a Fourier transform-based local texture estimator to improve the recovery of high-frequency details, achieving more precise texture representation. Recently, LMF~\cite{he2024latent} introduced a novel approach that decouples the decoding process using latent modulation, which separates shared latent decoding across the LR-HD (high-dimensional) space while maintaining competitive reconstruction quality. Despite significant progress, existing ASISR methods often struggle in real-world applications due to their high computational and storage demands. This limitation underscores the need for more computationally efficient approaches for ASISR.

\begin{figure*}[!t] 
\centering
\begin{minipage}{1.0\linewidth}
\centerline{\includegraphics[width=0.98\textwidth, height=0.32\textheight]{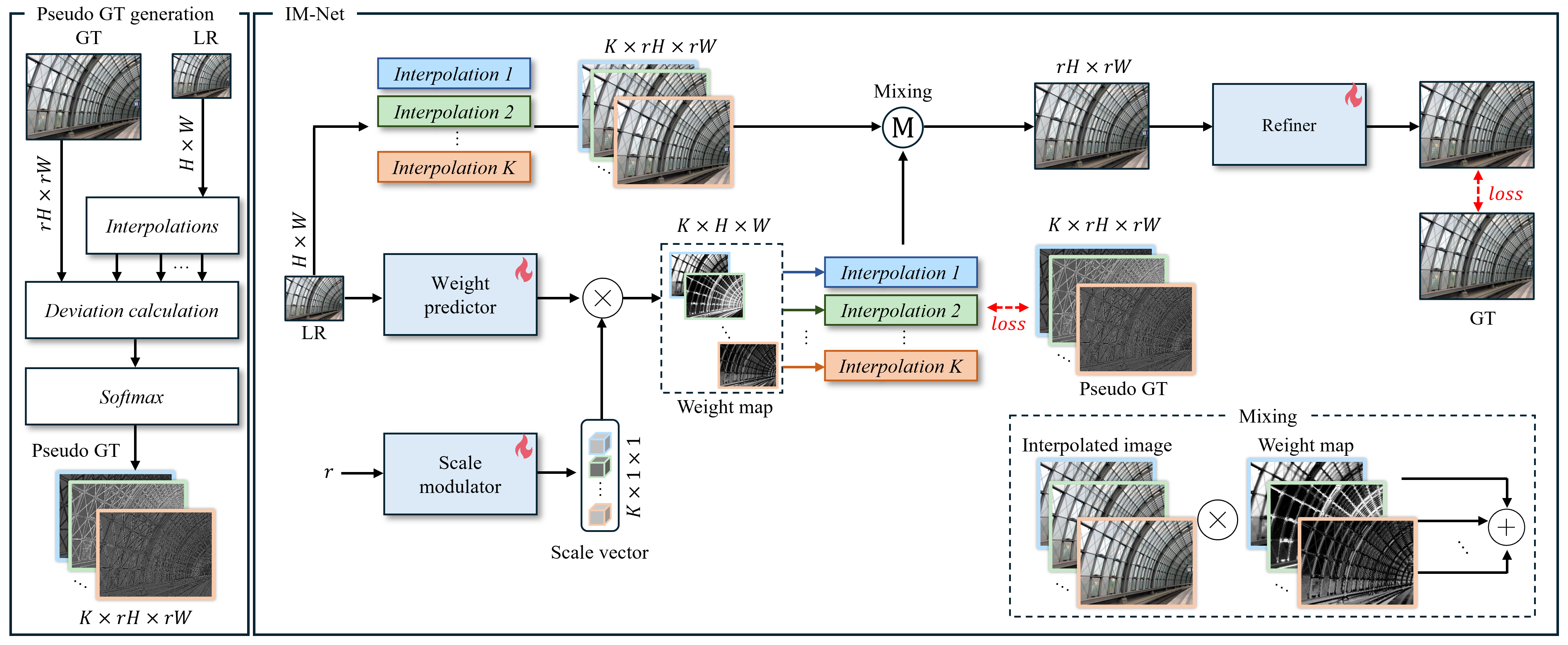}}
\end{minipage}
\caption{\textbf{Overall framework of IM-Net}. The input image is upsampled using $K$ different interpolation methods, and the interpolated images are combined through weighted averaging based on the weight maps predicted by the weight predictor and scale modulator. The refiner then post-processes the result to minimize differences from the GT image. As shown on the left, the pseudo GT weight maps are generated by comparing multi-interpolation results with the GT image to supervise the weight maps during the training of IM-Net.}
\label{fig_framework}
\end{figure*}

\section{Method}

\subsection{Motivation}
\cref{fig_motivation} compares the proposed IM-LUT with other SR approaches. (a) \textbf{Fixed-scale LUT-based SR}: Conventional LUT-based SR methods~\cite{jo2021practical, li2022mulut, li2024look} construct LUTs for a fixed set of scaling factors. Since LUTs are trained independently for a specific scale, a new set of LUTs must be generated for every desired scaling factor. This rigid approach significantly limits the applicability to ASISR. (b) \textbf{INR-based ASISR}: ASISR methods often utilize INRs to model images as continuous functions, enabling flexible resampling at arbitrary scales~\cite{chen2021learning, lee2022local, he2024latent}. However, this flexibility comes at the cost of high computational demands and increased storage requirements, limiting practical deployment. (c) \textbf{LUT-based single-kernel interpolation}: Recent methods~\cite{li2023learning, li2024lerf} leverage LUTs to adaptively apply interpolation, enabling ASISR. However, using a single class of interpolation functions uniformly across all regions makes it difficult to capture diverse local patterns and scale variations. Since different textures, edges, and smooth areas require distinct interpolation strategies, a single-kernel design lacks the flexibility needed across varying image structures and scaling factors. (d) \textbf{IM-LUT}: We propose a content-adaptive mixing of multiple interpolation kernels, significantly enhancing representation ability. By dynamically blending multiple interpolation kernels based on local image features and scale factors, our method achieves a richer and more expressive representation. This approach effectively adapts to diverse image structures, providing robust and adaptive SR across arbitrary scales while maintaining the computational efficiency of LUT-based approaches.

\subsection{Overview}
The proposed framework consists of two components: IM-Net and IM-LUT. IM-Net learns a pixel-wise interpolation mixing strategy based on local texture features and the target scale factor. Once IM-Net is trained, all learned components are mapped to LUTs, forming IM-LUT. As a result, IM-LUT enables network-free ASISR at the inference stage. The following subsections detail the components of IM-Net and the process of transferring IM-Net to IM-LUT.

\begin{figure}[!t] 
\centering
\begin{minipage}{1.0\linewidth}
\centerline{\includegraphics[width=0.93\textwidth, height=0.32\textheight]{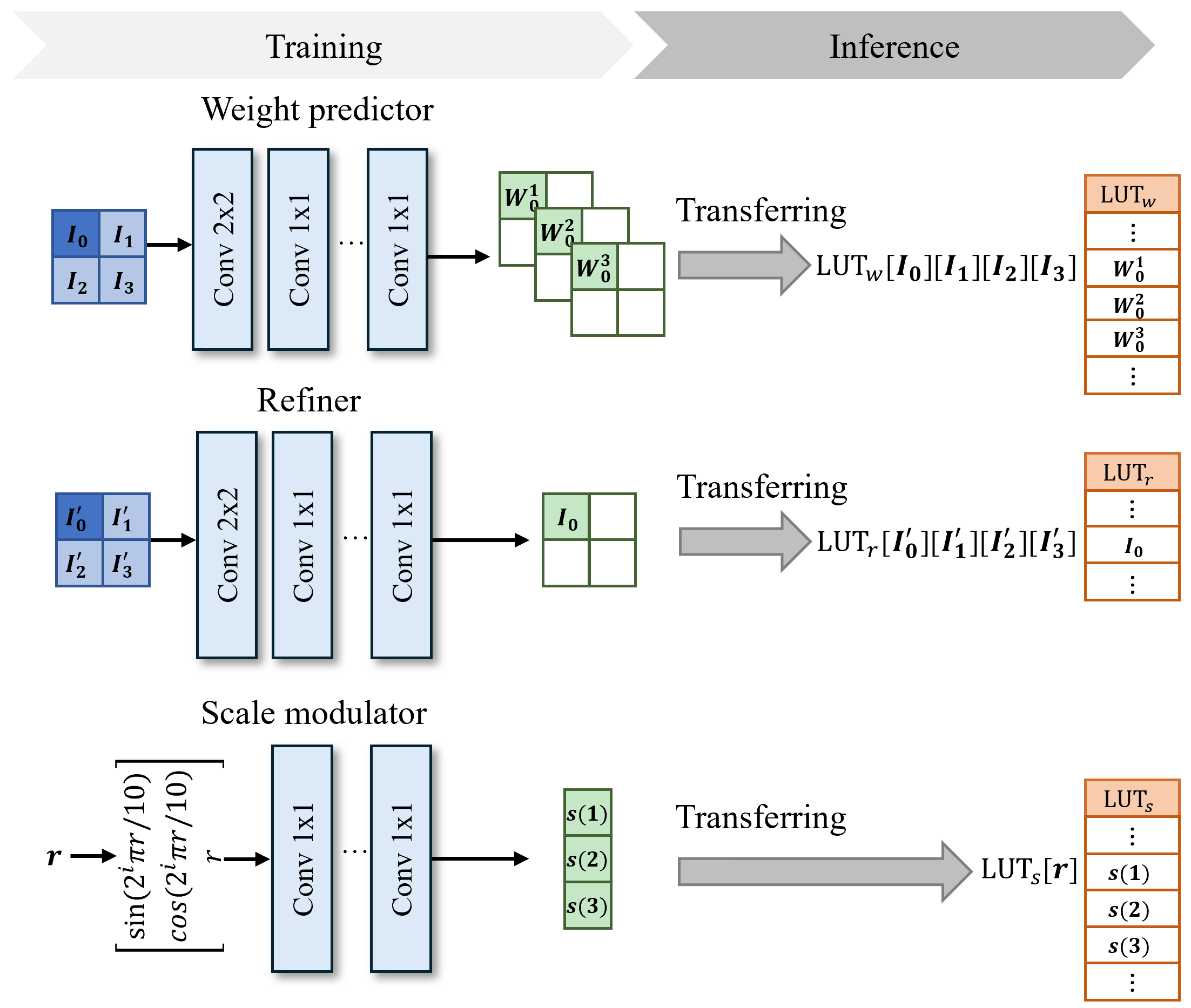}}
\end{minipage}
\caption{Illustration of the transfer from the neural network components in IM-Net to their corresponding LUT components in IM-LUT for $K$ = 3.}
\label{fig_transfer}
\end{figure}

\subsection{IM-Net Structure}
As shown in \cref{fig_framework}, IM-Net consists of three key components: a weight predictor $f_w(\cdot)$, a scale modulator $f_s(\cdot)$, and a refiner $f_r(\cdot)$. Given an LR image $ \bm{I}_{LR} \in \mathbb{R}^{H \times W}$, $K$ interpolation methods are applied to generate $K$ SR results with an upsampling factor $r$, forming $\mathcal{I}_{SR} = \{\bm{I}_{SR}^1, \bm{I}_{SR}^2, \cdots, \bm{I}_{SR}^K\}$, where $\bm{I}_{SR}^k \in \mathbb{R}^{rH \times rW}$. Our method is not limited to specific interpolation techniques; it is compatible with any efficient interpolation method, such as bilinear and bicubic interpolation. The results obtained using different combinations of interpolation methods will be discussed in~\cref{sec:experiments}.

Given $\bm{I}_{LR} $, the weight predictor estimates pixel-wise mixing weights for the $K$ interpolation methods, as follows: 
\begin{equation}
\mathcal{W} = f_w\left( \bm{I}_{LR}  \right),
\end{equation}
where $\mathcal{W} = \{\bm{W}^1, \bm{W}^2, \cdots, \bm{W}^K\}$ and $\bm{W}^k \in \mathbb{R}^{H \times W}$. In other words, the weight predictor operates in the LR domain, estimating pixel-wise mixing weights for the $K$ interpolation methods. Notably, the weight predictor is content-adaptive, as it assigns distinct weights to different interpolation methods based on local image characteristics within its receptive field.

Additionally, the target scale factor $r$ is fed into the scale modulator to produce a scale vector $\bm{s} \in \mathbb{R}^{K}$ as follows: 
\begin{equation}
\bm{s} = f_s\left(r\right). 
\end{equation}
Inspired by~\cite{wang2023deep}, the scale factor is first encoded using sinusoidal functions as follows:
\begin{equation}\label{eq:scale}
    \bm{r} = 
    \begin{bmatrix} 
        \sin(2^i \pi r / 10) \\
        \cos(2^i \pi r / 10) \\
        r
    \end{bmatrix}, \quad i = 0,1,\cdots,N,
\end{equation}
where $\bm{r}$ denotes the encoded scale vector, and $N$ determines its dimension. The encoded scale vector is then processed by linear layers, producing $\bm{s}$. The predicted weight maps and the scale vector are then multiplied, resulting in $\hat{\mathcal{W}} = \{{\bm{s}}(1)\bm{W}^1, \bm{s}(2)\bm{W}^2, \cdots, \bm{s}(K)\bm{W}^K\}$. As a result, the weight maps become scale-adaptive, dynamically modulated by the weights derived from the target scale.

Next, each scale-modulated weight map in $\hat{\mathcal{W}}$ is upsampled by its corresponding interpolation method, generating a set of HR weight maps, denoted as $\mathcal{W}_{SR} = \{\bm{W}_{SR}^1, \bm{W}_{SR}^2, \cdots, \bm{W}_{SR}^K\}$, where $\bm{W}_{SR}^k \in \mathbb{R}^{rH \times rW}$. The upsampled images and weights are then combined to yield an SR image ${\bm{I}}'_{SR}$ as 
\begin{equation}\label{eq:mixing}
{\bm{I}}'_{SR} = \bm{I}_{SR}^1 \odot \bm{W}_{SR}^1+ \cdots + \bm{I}_{SR}^K \odot \bm{W}_{SR}^K, \
\end{equation}
where $\odot$ represents element-wise multiplication.

Finally, the mixed image is processed by the refiner which mitigates residual distortions and enhances high-frequency details, producing a refined SR image ${\bm{I}}_{SR}$ as follows: 
\begin{equation}
{\bm{I}}_{SR} = f_r({\bm{I}}'_{SR}).
\end{equation}

The network structures of the weight predictor, scale modulator, and refiner are shown in~\cref{fig_transfer}. For clarity, the weight predictor and refiner are depicted as a single network structure with a receptive field of 2$\times$2; however, in our implementation, we employ three networks for each, leveraging the hierarchical and complementary indexing structures of MuLUT~\cite{li2024toward} along with the rotation ensemble strategy~\cite{jo2021practical}, resulting in a receptive field size of 5$\times$5.

\begin{figure}[!t] 
\centering
\begin{minipage}{1.0\linewidth}
\centerline{\includegraphics[width=0.98\textwidth, height=0.31\textheight]{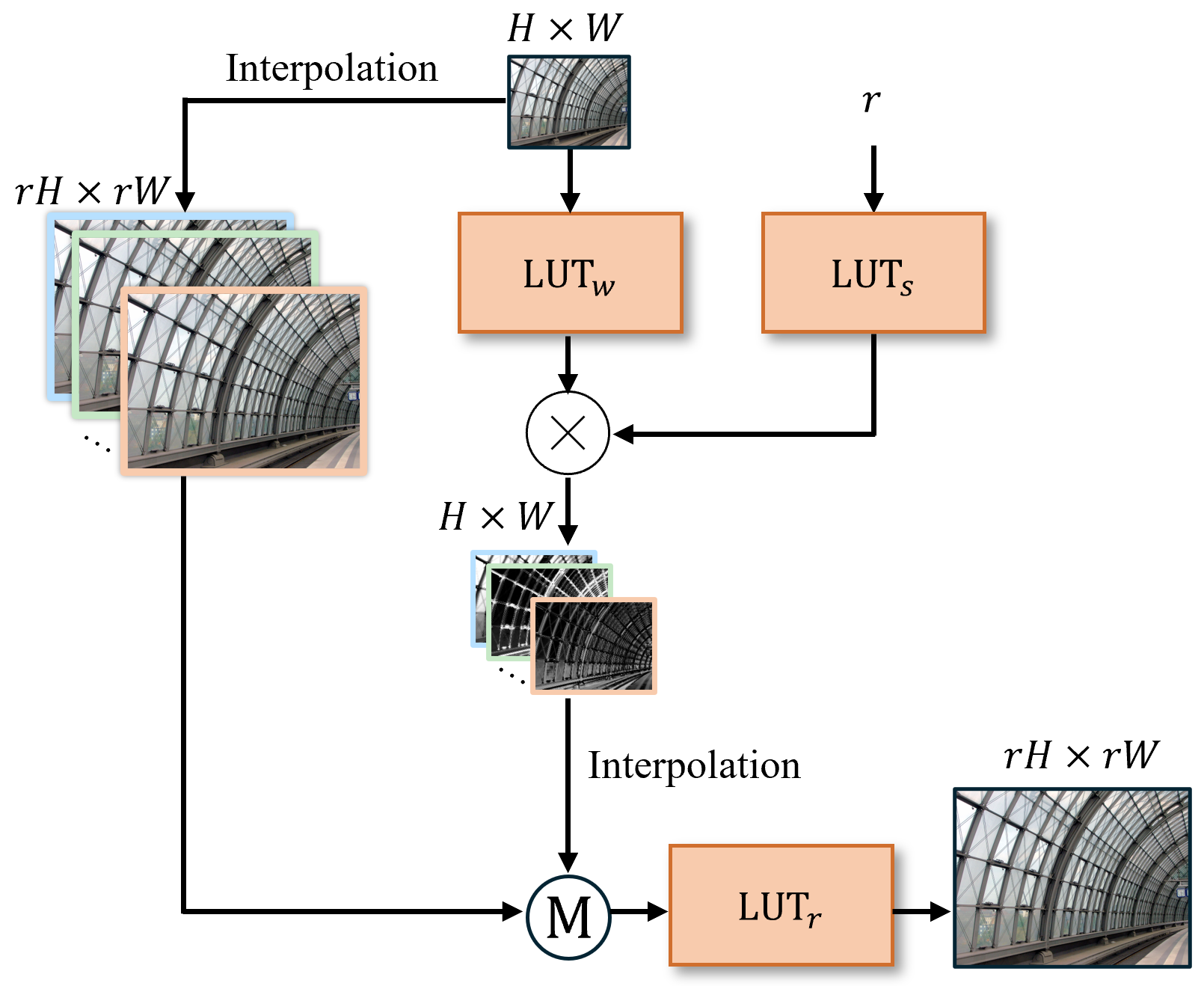}}
\end{minipage}
\caption{Illustration of IM-LUT at the inference stage.}
\label{fig_inference_}
\end{figure}

\subsection{IM-Net Training}

For the training of IM-Net, the standard reconstruction loss function, $\mathcal{L}_{\text{rec}}$, is calculated as follows:
\begin{equation}
\mathcal{L}_{\text{rec}} = \sum_{i,j} \left\| \bm{I}_{SR}(i,j) - \bm{I}_{GT}(i,j) \right\|_2,
\end{equation}
where $\bm{I}_{GT}$ represents the ground-truth (GT) HR image, and $\left\| \cdot \right\|_2$ represents the L2 norm. 

\begin{figure}[t] 
\centering
\begin{minipage}{1.0\linewidth}
\centerline{\includegraphics[width=0.98\textwidth, height=0.22\textheight]{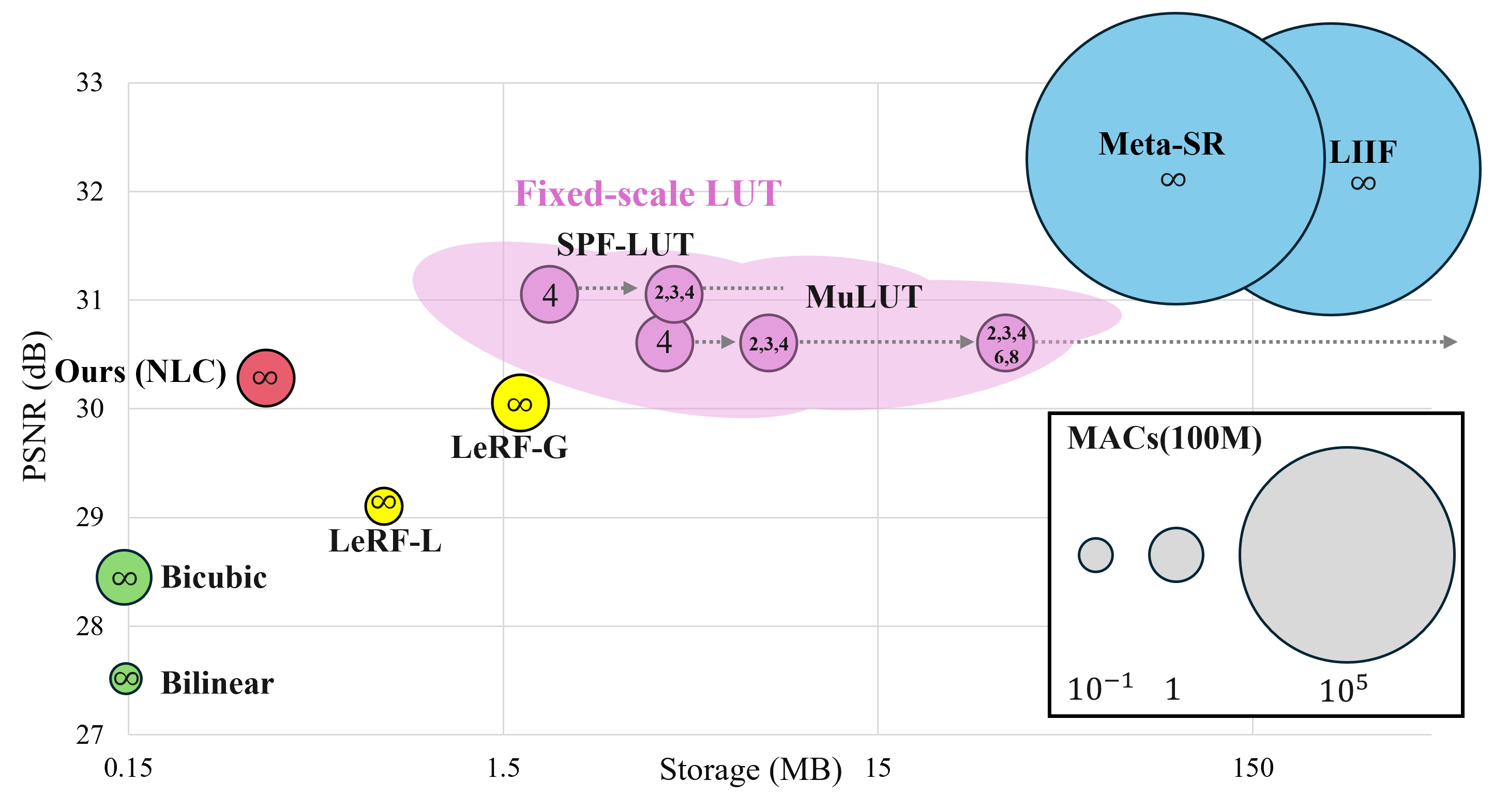}} 
\end{minipage}
\caption{Comparison in PSNR, MACs, and storage requirements for $\times$4 SR for producing a 1280$\times$720 image on Set 5. Numbers in circles denote supported upscaling factors(e.g., $\times$4, $\times\infty$).}
\label{fig_tradeoff}
\end{figure}
In addition to the reconstruction loss, we introduce a novel loss function, denoted as $\mathcal{L}_{\text{guide}}$, which guides the prediction of the weight maps. To effectively supervise the weight maps, we derive the pseudo GT weight maps by comparing the $K$ interpolation results, $\{\bm{I}_{SR}^1, \bm{I}_{SR}^2, \cdots, \bm{I}_{SR}^K\}$, with the GT image $\bm{I}_{GT}$. Specifically, the deviation from the GT image is computed for each interpolation method at every pixel as follows:
\begin{equation}
E^k(i, j) = \left|\bm{I}^{k}_{SR}(i, j)- \bm{I}_{GT}(i, j)\right|, k = 1,2,\cdots,K,\
\end{equation}
where $E^k$ represents the deviation map for the $k$-th interpolation method, and $\left| \cdot \right|$ measures the L1 norm. Then, we obtain a set of pseudo GT weight maps, denoted as $\mathcal{\overline{W}} = \{\bm{\overline{W}}^1, \bm{\overline{W}}^2, \dots, \bm{\overline{W}}^K\}$. Each $ \bm{\overline{W}}^k \in \mathbb{R}^{rH \times rW} $ is defined using a temperature softmax function as follows:
\begin{equation}\label{eq:weight}
\bm{\overline{W}}^k(i, j) = \frac{\exp(-\beta E^k(i, j))}{\sum_{m=1}^{K} \exp(-\beta E^m(i, j))},
\end{equation}
where \( \beta \) is a positive constant that controls the sharpness of weight distribution. Intuitively, $\bm{\overline{W}}^k(i, j)$ assigns higher values to interpolation methods whose results $\bm{I}^{k}_{SR}(i, j)$ are closer to $\bm{I}_{GT}(i, j)$.
During training, we utilize a mixed-scale setting~\cite{li2023learning}, ensuring that pseudo GT weight maps vary across different target scale factors. Finally, the weight guidance loss $\mathcal{L}_{\text{guide}}$ is computed as follows:
\begin{equation}
\mathcal{L}_{\text{guide}} = \sum_{k} \sum_{i,j}  \left| \left| \bm{W}_{SR}^k(i,j) - \bm{\overline{W}}^{k}(i,j) \right| \right|_2.
\label{eq:loss_guide}
\end{equation}
The total loss function for training IM-Net is given as follows:
\begin{equation}
\mathcal{L}_{total} := \mathcal{L}_{rec} + \lambda\mathcal{L}_{guide},
\label{eq:loss_total}
\end{equation}
where $\lambda$ is a hyperparameter that balances the two loss terms.

\begin{table}[t]
\centering
\renewcommand{\arraystretch}{1.0}
\setlength{\tabcolsep}{2.0pt}
\begin{tabular}{lccccc}
\toprule
\multirow{2}{*}{Method} & \multirow{2}{*}{$\text{MACs}_\downarrow$} & \multirow{2}{*}{$\text{Storage}_\downarrow$} & \multicolumn{3}{c}{PSNR (dB)} \\
\cmidrule(lr){4-6}
& & & {$\times2_\uparrow$} & {$\times3_\uparrow$} & {$\times4_\uparrow$} \\
\midrule
Nearest       & -           & -        & 30.84 & 27.91&26.25 \\
Bilinear      & 14.74M      & -        & 32.23 & 29.53&27.55\\
Bicubic       & 51.61M      & -        & 33.64 & 30.39&28.42\\
Lanczos2      & 110.59M     & -        & 33.70 & 30.41&28.44 \\
\midrule

MuLUT*~\cite{li2024toward}      & 86.40M      & 1.20MB   & 36.65 & 33.25 & 30.60 \\
SPF-LUT*~\cite{li2024look}       & 402.74M      & 934.90KB   & 36.61 & 33.33 & 31.05 \\
\midrule
LeRF-L~\cite{li2024lerf}        & 53.11M      & 737.28KB    & 34.77 & 30.73 & 29.10 \\
LeRF-G~\cite{li2023learning}        & 110.12M      & 1.67MB   & 35.62 & 31.95 & 30.05 \\
\rowcolor{orange!20}
Ours (NL)       & 104.81M     & 322.33KB    & 35.83 & 31.94 & 30.07\\
\rowcolor{orange!20}
Ours (NLC)       & 225.76M     & 360.93KB    & 36.20 & 32.40 &30.19\\
\rowcolor{orange!20}
Ours (NLCZ)       & 463.76M     & 399.21KB    & 36.35 & 32.51 &30.32\\
\midrule
MetaSR~\cite{hu2019meta}        & 5.47T    & 85.59MB  & 38.12 &  34.71 & 32.48 \\
LIIF~\cite{chen2021learning}    & 6.34T    & 255.76MB & 38.08 &  34.59 & 32.37 \\
\bottomrule
\end{tabular}
\caption{PSNR comparison of SR methods in terms of MACs and storage requirements for $\times$2 SR for producing a 1280$\times$720 image and PSNR for $\times$2, $\times$3, and $\times$4 SR on the Set 5 dataset. *denotes fixed-scale LUT methods.}
\label{tab:efficiency}
\end{table}

\subsection{IM-Net to IM-LUT}
After training IM-Net, its three neural network components, \textit{i.e.}, the weight predictor, scale modulator, and refiner, are transferred to their LUT counterparts, denoted as $\text{LUT}_{w}$, $\text{LUT}_{s}$, and $\text{LUT}_{r}$, respectively, as shown in~\cref{fig_transfer}. This transfer is performed by computing output values of the learned networks for input values and storing them in their corresponding LUTs. For example, $\text{LUT}_{w}$ is implemented as a 4D LUT, where four input pixel values serve as the index, and their corresponding weights, \textit{i.e.}, $K$ weight values for the $K$ interpolation methods, are stored at the indexed location. $\text{LUT}_{s}$ and $\text{LUT}_{r}$ are defined as a 1D LUT and a 4D LUT, respectively, in a similar manner. 

To reduce the storage requirements of the LUTs, we adopt the strategy~\cite{jo2021practical} that applies uniform sampling to the input values. Specifically, input pixel values for $\text{LUT}_{w}$ and $\text{LUT}_{s}$ are quantized to 9 bins and 7 bins, respectively, while input scale values for $\text{LUT}_{r}$ are quantized to 17 bins. Experimental results with different sampling intervals can be found in the supplementary material. After the LUT transfer, the LUTs are refined by applying the fine-tuning strategy following~\cite{li2024toward}.
\begin{table*}[t]
    \centering
    \renewcommand{\arraystretch}{1.0}
    \setlength{\tabcolsep}{2.3pt}
    \begin{tabular}{l|cccc|cccc|cccc|cccc}
    \toprule
        \multirow{2}{*}{Method} & \multicolumn{4}{c|}{Set5} & \multicolumn{4}{c|}{Set14} & \multicolumn{4}{c|}{BSDS100} & \multicolumn{4}{c}{Urban100} \\
        & $\frac{\times 1.5}{\times 1.5}$ & $\frac{\times 1.5}{\times 2.0}$ & $\frac{\times 2.0}{\times 2.0}$ & $\frac{\times 2.0}{\times 2.4}$ 
        & $\frac{\times 1.5}{\times 1.5}$ & $\frac{\times 1.5}{\times 2.0}$ & $\frac{\times 2.0}{\times 2.0}$ & $\frac{\times 2.0}{\times 2.4}$ 
        & $\frac{\times 1.5}{\times 1.5}$ & $\frac{\times 1.5}{\times 2.0}$ & $\frac{\times 2.0}{\times 2.0}$ & $\frac{\times 2.0}{\times 2.4}$ 
        & $\frac{\times 1.5}{\times 1.5}$ & $\frac{\times 1.5}{\times 2.0}$ & $\frac{\times 2.0}{\times 2.0}$ & $\frac{\times 2.0}{\times 2.4}$ \\
        \midrule
        Nearest & 31.34 & 31.07 & 30.84 & 29.63 & 29.15 & 28.84 & 28.57 & 27.70 & 28.99 & 28.72 & 28.40 & 27.62 & 26.21 & 25.91 & 25.62 & 24.78 \\
        Bilinear & 34.99 & 33.19 & 32.23 & 31.49 & 31.68 & 30.26 & 29.24 & 28.70 & 30.92 & 29.66 & 28.67 & 28.20 & 28.24 & 26.91 & 25.96 & 25.46 \\
        Bicubic & 36.76 & 34.68 & 33.64 & 32.70 & 33.07 & 31.45 & 30.32 & 29.62 & 32.14 & 30.67 & 29.54 & 28.93 & 29.50 & 27.95 & 26.87 & 26.22 \\
        Lanczos3 & \textbf{37.61} & \textbf{35.31} & \textbf{34.23} & \textbf{33.24}& \textbf{33.75} & \textbf{31.97} & \textbf{30.76} & \textbf{30.02} & \textbf{32.74} & \textbf{31.11} & \textbf{29.89} & \textbf{29.23} & \textbf{30.12} & \textbf{28.42} & \textbf{27.25} & \textbf{26.55} \\
        \midrule
        LeRF-L & 37.70 & 35.76 & 34.77 & 33.57 & 33.92 & 32.33 & 31.16 & 30.26 & 33.17 &31.53 & 30.27 & 29.53 & 30.46 & 28.78 & 27.59 & 26.81 \\
        LeRF-G & 38.20 & 36.51 & 35.62 & 34.63 & \textbf{34.60} & 33.01 & 31.94 & 31.07 & \textbf{33.70} & 32.04 & 30.79 & 30.05 & \textbf{31.78} & 29.98 & 28.76 & 27.89 \\
        \rowcolor{orange!20}
        Ours (NLC) & \textbf{38.30} & \textbf{37.00} & \textbf{36.20} & \textbf{35.02} & {34.34} & \textbf{33.10} & \textbf{32.14} & \textbf{31.21} & {33.40} & \textbf{32.14} & \textbf{30.93} & \textbf{30.20} & 31.60 & \textbf{30.03} & \textbf{28.81} & \textbf{28.05} \\
        \midrule
        Meta-SR & \textbf{41.29} & - & \textbf{38.12} & - & \textbf{37.47} & - & \textbf{33.99} & - & \textbf{35.79} & - & \textbf{32.32} & - & 35.85 & - & \textbf{32.98} & -  \\
        LIIF & 41.22 & \textbf{38.99} & 38.08 & \textbf{36.99} & 37.44 & \textbf{35.31} & 33.96 & \textbf{32.95} & 35.75 & \textbf{33.68} & 32.28 & \textbf{31.45} & \textbf{36.70} & \textbf{34.08} & 32.84 & \textbf{31.70} \\
        \bottomrule
    \end{tabular}

    \vspace{5pt}

    \begin{tabular}{l|cccc|cccc|cccc|cccc}
        \toprule
        \multirow{2}{*}{Method} & \multicolumn{4}{c|}{Set5} & \multicolumn{4}{c|}{Set14} & \multicolumn{4}{c|}{BSDS100} & \multicolumn{4}{c}{Urban100} \\
        & $\frac{\times 2.0}{\times 3.0}$ & $\frac{\times 3.0}{\times 3.0}$ & $\frac{\times 3.0}{\times 4.0}$ & $\frac{\times 4.0}{\times 4.0}$ 
        & $\frac{\times 2.0}{\times 3.0}$ & $\frac{\times 3.0}{\times 3.0}$ & $\frac{\times 3.0}{\times 4.0}$ & $\frac{\times 4.0}{\times 4.0}$ 
        & $\frac{\times 2.0}{\times 3.0}$ & $\frac{\times 3.0}{\times 3.0}$ & $\frac{\times 3.0}{\times 4.0}$ & $\frac{\times 4.0}{\times 4.0}$ 
        & $\frac{\times 2.0}{\times 3.0}$ & $\frac{\times 3.0}{\times 3.0}$ & $\frac{\times 3.0}{\times 4.0}$ & $\frac{\times 4.0}{\times 4.0}$ \\
        \midrule
        Nearest & 28.87 & 27.91 & 26.88 & 26.25 & 27.07 & 26.08 & 25.33 & 24.74 & 27.12 & 26.17 & 25.57 & 25.03 & 24.25 & 23.34 & 22.68 & 22.17 \\
        Bilinear & 30.43 & 29.53 & 28.27 & 27.55 & 27.94 & 27.04 & 26.16 & 25.51 & 27.60 & 26.77 & 26.11 & 25.53 & 24.81 & 23.99 & 23.26 & 22.68 \\
        Bicubic & 31.41 & 30.39 & 29.12 & 28.42 & 28.70 & 27.63 & 26.75 & 26.09 & 28.18 & 27.20 & 26.53 & 25.95 & 25.43 & 24.45 & 23.71 & 23.14 \\
        Lanczos3 & \textbf{31.85} & \textbf{30.79} & \textbf{29.49} & \textbf{28.78}& \textbf{29.04} & \textbf{27.91} & \textbf{27.01} & \textbf{26.31} & \textbf{28.43} & \textbf{27.39} & \textbf{26.70} & \textbf{26.10} & \textbf{25.71} & \textbf{24.68} & \textbf{23.92} & \textbf{23.32} \\
        \midrule
        LeRF-L & 31.93 & 30.73 & 29.71 & 29.10 & 29.13 & 27.88 & 27.13 & 26.52 & 28.62 & 27.43 & 26.83 & 26.29 & 25.84 & 24.68 & 24.04 & 23.50 \\
        LeRF-G & 33.08 & 31.95 & 30.77 & 30.05 & 30.01 & 28.79 & 27.99 & 27.28 & 29.11 & 27.97 & 27.28 & 26.66 & 26.83 & 25.62 & 24.81 & 24.16 \\
        \rowcolor{orange!20}
        Ours (NLC) & \textbf{33.63} & \textbf{32.40} & \textbf{31.00} & \textbf{30.19} & \textbf{30.31} & \textbf{29.11} & \textbf{28.11} & \textbf{27.33} & \textbf{29.25} & \textbf{28.08} & \textbf{27.34} & \textbf{26.71} & \textbf{26.97} & \textbf{25.78} & \textbf{24.83} & \textbf{24.18} \\
        \midrule
        Meta-SR & - & \textbf{34.71} & - & \textbf{32.48} & - & \textbf{30.56} & - & \textbf{28.83} & - & \textbf{29.26} & - & \textbf{27.73} & - & \textbf{28.91} & - & \textbf{26.69}  \\
        LIIF & \textbf{35.63} & 34.59 & \textbf{33.17} & 32.37 & \textbf{31.68} & 30.39 & \textbf{29.45} & \textbf{28.65} & \textbf{30.47} & 29.24 & \textbf{28.42} & \textbf{27.73} & \textbf{30.23} & 28.80 & \textbf{27.55} & 26.66 \\
        \bottomrule
    \end{tabular}
    \caption{Performance comparison of ASISR methods for various target scale factors in terms of PSNR. $\frac{\times r_h}{\times r_w}$ represents upsampling $r_h$ times along the height and $r_w$ times along the width. The first four rows in each table correspond to traditional interpolation methods, the three rows of the middle represent LUT-based methods, and the last two rows correspond to neural network-based methods. The highest scores in each category are boldfaced.}
    \label{tab:psnr}
\end{table*}

At the inference stage, an SR image can be reconstructed without requiring any neural network components, as shown in ~\cref{fig_inference_}. The pixel values of the input LR image and the target scale factor are used to index \( \text{LUT}_w \) and \( \text{LUT}_s \), respectively. For non-sampled points in \( \text{LUT}_w \) and \( \text{LUT}_r \), tetrahedral interpolation is performed using the values of their nearest sampled points~\cite{jo2021practical}. In contrast, the nearest scale value in \( \text{LUT}_s \) is directly applied. After reconstructing the weight maps using \( \text{LUT}_w \) and \( \text{LUT}_s \), the \( K \) interpolation methods are applied to obtain HR weight maps, which are then mixed with the $K$ interpolated images using \eqref{eq:mixing}. Finally, \( \text{LUT}_r \) is applied to refine the pixel values, producing the SR image. The rotation ensemble strategy~\cite{jo2021practical} is also employed at the inference stage, effectively expanding the receptive field without increasing the storage requirements of LUTs.

\section{Experiments}\label{sec:experiments}
\subsection{Experimental Settings}
\noindent \textbf{Datasets and Metrics.} We train IM-Net using the DIV2K dataset~\cite{agustsson2017ntire} in a mixed-scale setting~\cite{li2023learning}, ensuring that the model effectively generalizes across various scale factors. For evaluation, we use multiple benchmark datasets, including Set5, Set14, BSDS100~\cite{martin2001database}, and Urban100~\cite{huang2015single}. To assess SR performance, we use the peak signal-to-noise ratio (PSNR). Additionally, we evaluate computational efficiency in terms of theoretical multiply-accumulate (MAC) operations, storage size and running time.

\noindent \textbf{Implementation Details.}
The IM-Net is trained for a total of 100k iterations with a batch size of 16 and an initial learning rate of \( 1e^{-3} \). The dimension $N$ in \eqref{eq:scale} and the weighting factor \( \lambda \) in \eqref{eq:loss_total} is set as 16 and 0.1, respectively. For producing pseudo GT weight maps, we set the temperature parameter \( \beta \) in \eqref{eq:weight} as 0.1.

\begin{figure*}[!t] 
\centering
\begin{minipage}{1.0\linewidth}
\centerline{\includegraphics[width=0.98\textwidth, height=0.35\textheight]{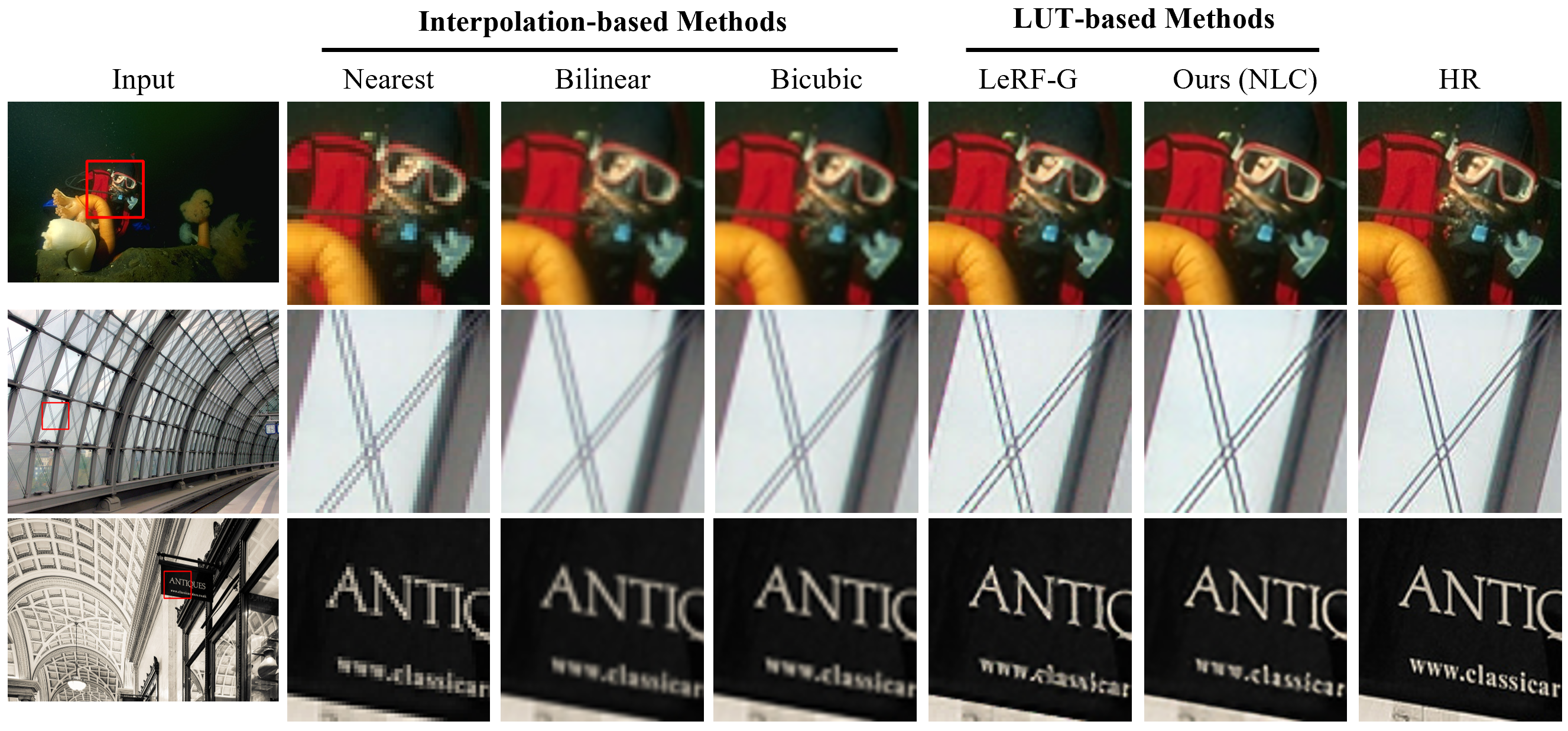}}
\end{minipage}
\caption{Comparison of $\times 2$ SR results of interpolation-based methods and LUT-based methods.}  

\label{fig_sr}
\end{figure*}

\noindent
\subsection{Performance Comparison} 
\cref{tab:efficiency} presents performance comparisons of different SR methods in terms of PSNR, storage requirements, and computational efficiency. As shown in \cref{tab:efficiency}, previous LUT-based ASISR methods~\cite{li2023learning, li2024lerf} offer improved reconstruction quality over traditional interpolation techniques. In this table, the option with the minimal MACs is set as default choice for each available interpolation combination: \textit{e.g.}, NLCZ (Nearest neighbor, biLinear, biCubic, lancZos interpolation). Specifically, we achieve a significant improvement in reconstruction quality while also substantially reducing storage requirements. Fixed-scale LUT-based SR methods trained separately for each corresponding scale ~\cite{li2024toward, li2024look} achieve higher PSNR scores than ours; however, they are not applicable to ASISR. In addition, IM-LUT requires significantly less memory than these methods. Lastly, although neural network-based ASISR methods~\cite{hu2019meta, chen2021learning} achieve the highest PSNR scores, they demand substantial storage and require tera-scale MACs, making them impractical for resource-constrained environments. \cref{fig_tradeoff} further illustrates the performance-efficiency trade-off, where storage size and MACs are plotted on a logarithmic scale. IM-LUT (NLC) achieves a favorable balance between efficiency and reconstruction quality compared to all previous approaches. Specifically, compared to previous LUT-based ASISR methods~\cite{li2023learning, li2024lerf}, IM-LUT achieves higher PSNR while maintaining similar MAC efficiency with significantly lower storage requirements.

\cref{tab:psnr} compares SR methods for various scale factors, including non-integer and anisotropic scaling. Our approach, \textit{i.e.}, IM-LUT (NLC), significantly outperforms traditional interpolation methods. Moreover, compared to  LeRF~\cite{li2023learning, li2024lerf}, the state-of-the-art LUT-based ASISR method, our method achieves higher PSNR scores across multiple scale factors. Additionally, we compare our method with representative neural network-based ASISR approaches~\cite{hu2019meta,chen2021learning}, which attain higher PSNR scores than ours but at the cost of substantial computational overhead.

\begin{table}[t]
\centering

\renewcommand{\arraystretch}{1.0}
\setlength{\tabcolsep}{2.0pt}
\begin{tabular}{lccccc}
\toprule
{Method} & {$\text{CPU Runtime}_\downarrow$} & {$\text{MACs}_\downarrow$} & {$\text{Storage}_\downarrow$} & {$\times2_\uparrow$}\\
\midrule
Nearest        & 1.70ms & --       & --          & 30.84 \\
Bilinear       & 2.26ms & 14.74M   & --          & 32.23 \\
Bicubic        & 3.97ms & 51.61M   & --          & 33.64 \\
\midrule
LeRF-G~\cite{li2023learning}    & 5.51s & 110.12M  & 1.67MB     & 35.62 \\
\rowcolor{orange!20}
Ours (NLC)     & 2.73s & 225.76M  & 360.93KB     & 36.20 \\
\midrule
LIIF~\cite{chen2021learning}  & 48.33s & 6.34T    & 255.76MB     & 38.08 \\
\bottomrule
\end{tabular}

\caption{Performance comparison of ASISR methods in terms of runtime, MACs and storage requirements for $\times2$ SR for producing a 1280$\times$720 image and PSNR for $\times2$ SR on the Set5 dataset.}
\label{tab:runtime}
\end{table}

Furthermore, in \cref{tab:runtime}, we evaluate the runtime of ASISR methods following~\cite{li2023learning} on our desktop computer. All methods perform $\times2$ SR on AMD Ryzen 9 3950X 16-Core Processor 3.5GHz CPU to produce a 1280$\times$720 image with the results averaged across 10 trials. LeRF-G~\cite{li2023learning}, which employs parallel and cascading LUT structures~\cite{li2022mulut}, require additional runtime and storage size, whereas our approach, which only uses a parallel structure, achieves faster processing by utilizing a mixing strategy with basic interpolations.

Finally, \cref{fig_sr} compares SR results from interpolation methods, LeRF~\cite{li2023learning}, and our method. From left to right, the reconstructed images exhibit progressive improvements in sharpness and detail preservation. Traditional interpolation methods tend to produce overly smooth results, failing to recover fine details. LeRF suffers from ringing artifacts and blurriness due to its reliance on a single function. In contrast, IM-LUT effectively mitigates these issues by leveraging interpolation mixing, by producing more accurate textures. Our method provides a favorable balance between detail preservation and computational efficiency, demonstrating its effectiveness in generating high-quality SR images. More visual comparisons of the SR results are provided in the supplementary material.

\noindent \subsection{Analysis of Weight Maps} 
\begin{figure}[!t] 
\centering
\begin{minipage}{1.0\linewidth}
\centerline{\includegraphics[width=0.98\textwidth, height=0.31\textheight]{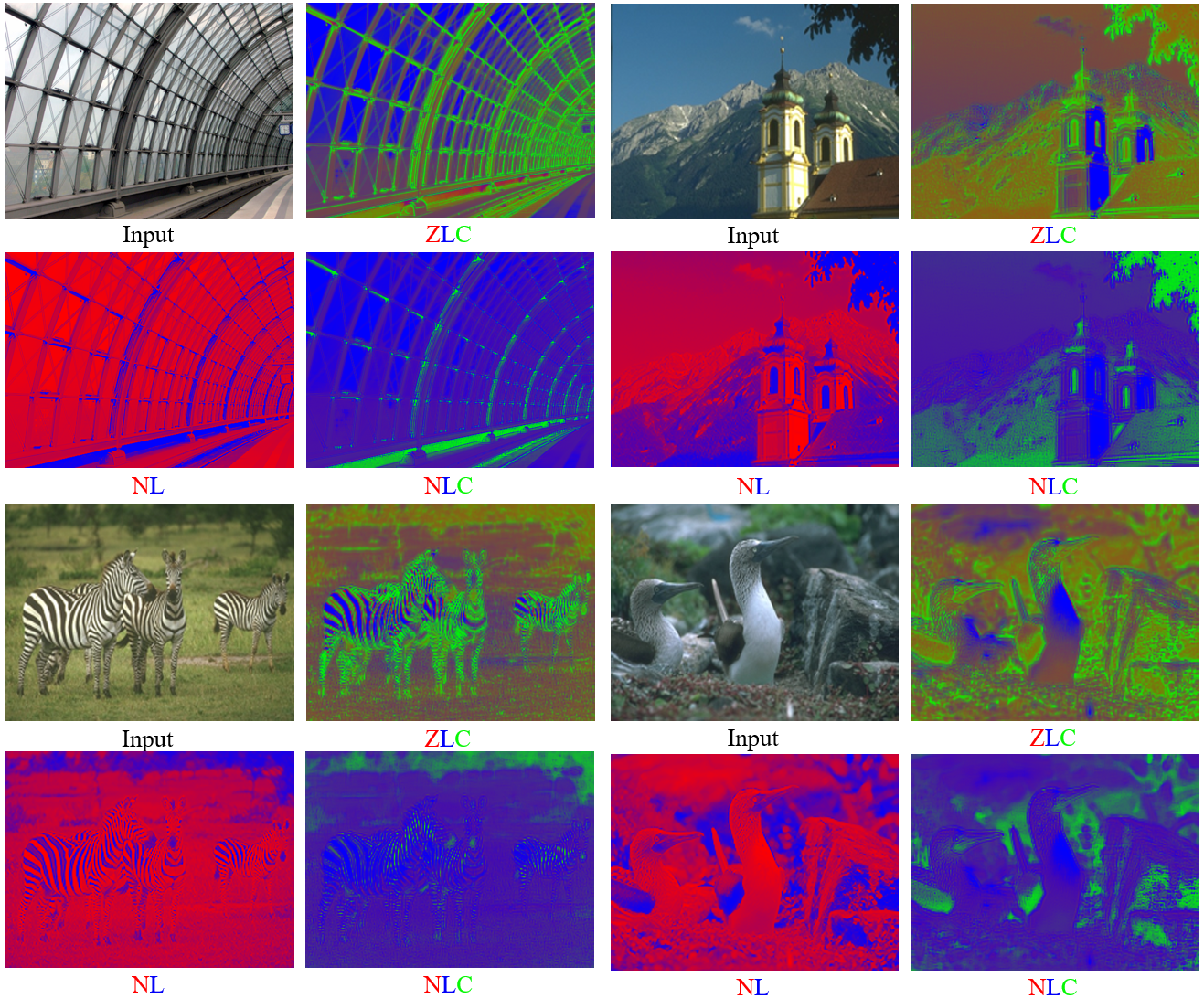}}
\end{minipage}
\caption{Visualization of predicted weight maps for different combinations of interpolations. The colored maps illustrate the assigned interpolation weights, where each interpolation method (Nearest neighbor, biLinear, biCubic, lancZos) is represented by a distinct color. Pure colors indicate dominant contributions, while mixed colors reflect blended influences.}  

\label{fig_visualization}
\end{figure}

To better understand the behavior of the proposed interpolation mixing, \cref{fig_visualization} illustrates interpolation weight maps for various interpolation combinations. These visualizations show that our method dynamically adapts interpolation weights. More importantly, our method does not merely select a single interpolation method over others; rather, it blends them adaptively based on spatial context. For example, in regions with moderately high-frequency textures, \textit{e.g.}, repetitive structures such as railings, bicubic and Lanczos2 contribute predominantly but remain softly mixed with bilinear interpolation to mitigate excessive ringing artifacts. Similarly, in smoother areas, bilinear interpolation does not fully dominate; instead, a subtle yet significant contribution from more complex interpolation methods ensures a balance between over-smoothing and detail preservation. These results highlight that the contribution of each interpolation method varies depending on the specific combination used, with their interplay being context-dependent rather than static, emphasizing the importance of adaptive mixing strategies.

\noindent \subsection{Ablation Study}
\cref{tab:ablation} presents the results of an ablation study on different components of IM-Net. The results demonstrate that incorporating the guidance loss, scale modulator, and refiner consistently improves reconstruction quality across different upsampling factors. In particular, the combination of all three components leads to the best overall performance, demonstrating their complementary contributions to SR quality. Additionally, the performance-efficiency trade-off for different interpolation combinations is provided in the supplementary  material.

\begin{table}[t]
    \centering
    \begin{tabular}{c c c c | c c c}
        \toprule
        \multicolumn{4}{c|}{{Method}} & \multicolumn{3}{c}{{PSNR}} \\
        \midrule
        $\mathcal{L}_{rec}$ & $\mathcal{L}_{guide}$ & $f_s(\cdot)$ & $f_r(\cdot)$ & $\times$2 & $\times$3 & $\times$4 \\
        \midrule
        \checkmark &  &  &  & 35.81 & 31.95 & 29.85 \\
        \checkmark & \checkmark &  &  & 35.95 & 32.02 & 29.90 \\
        \checkmark &  & \checkmark &  & 36.04 & 32.13 & 29.98 \\
        \checkmark &  &  & \checkmark & 36.26 & 32.53 & 30.39 \\
        \checkmark & \checkmark & \checkmark &  & 36.11 & 32.20 & 30.03 \\
        \checkmark & \checkmark &  & \checkmark & 36.40 & 32.58 & 30.43 \\
        \checkmark &  & \checkmark & \checkmark & 36.37 & 32.62 & 30.40 \\
        \rowcolor{orange!20}
        \checkmark & \checkmark & \checkmark & \checkmark & 36.48 & 32.64 & 30.46 \\
        \bottomrule
    \end{tabular}
    \caption{Ablation study on losses, scale modulator, and refiner for $\times$2, $\times$3, and $\times$4 SR on Set5.}
    \label{tab:ablation}
\end{table}

\subsection{Limitation}
Despite its efficiency and ability to operate without neural networks during inference, the performance of IM-LUT is inherently constrained by the interpolation methods it incorporates, leading to lower PSNR scores than state-of-the-art ASISR methods, particularly INR-based approaches~\cite{he2024latent, chen2021learning, lee2022local}. Future work could explore integrating more complementary interpolation methods to overcome this limitation.

\section{Conclusion}
In this paper, we introduced IM-LUT, a framework that combines the efficiency of LUT-based method with the scale flexibility of interpolation method. By leveraging multiple interpolation methods and dynamically assigning pixel-wise weights, IM-LUT effectively adapts to varying textures and target scale factors while maintaining computational efficiency across different scale factors. This interpolation mixing strategy enhances the representation ability of kernels, allowing the model to better capture complex patterns and scale variations. Through the transfer of IM-Net to IM-LUT, we achieve state-of-the-art performance among LUT-based ASISR methods while significantly reducing storage and computational costs compared to network-based ASISR approaches. Experimental results demonstrate that IM-LUT provides a favorable balance between reconstruction quality and efficiency, making it well-suited for real-world applications.

\section*{Acknowledgement}
{This work was supported by the National Research Foundation of Korea (NRF) grant funded by the Korea government (MSIT) (No. RS-2022-NR070077).}

{
    \small
    \bibliographystyle{ieeenat_fullname}
    \bibliography{main}
}

\newpage
\section{Additional Experiments}

\subsection{Effect of Sampling Intervals}

To investigate the impact of LUT sampling intervals on the weight predictor $f_w$, we conduct experiments NLC combination by fixing the refiner $f_r$ and varying the quantization levels of $f_w$. Table~\ref{tab:sampling} summarizes the results for $\times2$ super-resolution (SR) with LUT fine-tuning~\cite{li2024toward}, highlighting the relationship between LUT size and PSNR. 

As shown in Table~\ref{tab:sampling}, larger LUT sizes generally yield higher PSNR values, indicating that finer quantization enhances interpolation mixing precision and improves reconstruction quality. However, this comes at the cost of increased storage requirements, which can be impractical for memory-constrained scenarios.

To balance efficiency and quality, we adopt a sampling interval of $2^5$ (Ours NLC), which offers a favorable trade-off. While lower intervals (\textit{e.g.}, $2^0$, $2^2$) yield marginal PSNR improvements, they require significantly larger storage. Conversely, higher intervals (\textit{e.g.}, $2^6$) reduce storage demands but degrade image quality. Our setting minimizes storage overhead while maintaining PSNR close to that of the full LUT, demonstrating that moderate LUT quantization effectively preserves SR performance without excessive memory consumption.

\begin{table}[t]
    \centering
    \begin{tabular}{lcc}
        \toprule
        \textbf{Sampling} & \textbf{Size} & \textbf{PSNR}  \\
        \midrule
        $2^0$ (Full LUT)  & 73.13GB  & 36.48   \\
        $2^1$             & 4.64GB  & 36.46   \\
        $2^2$             & 306.67MB  &  36.43  \\
        $2^3$             & 20.59MB  &  36.41  \\
        $2^4$             & 1.67MB  &  36.34  \\
        $2^5$       & 360.02KB  & 36.20   \\
        $2^6$             &  255.67KB  & 35.58   \\
        
        \bottomrule
    \end{tabular}
\caption{Effect of lut sampling interval on storage size and PSNR for $\times 2$ SR on the Set5 dataset.}
\label{tab:sampling}
\end{table}

\subsection{Various Combinations of IM-LUT}

\begin{figure}[!t] 
\centering
\begin{minipage}{1.0\linewidth}
\centerline{\includegraphics[width=0.96\textwidth, height=0.3\textheight]{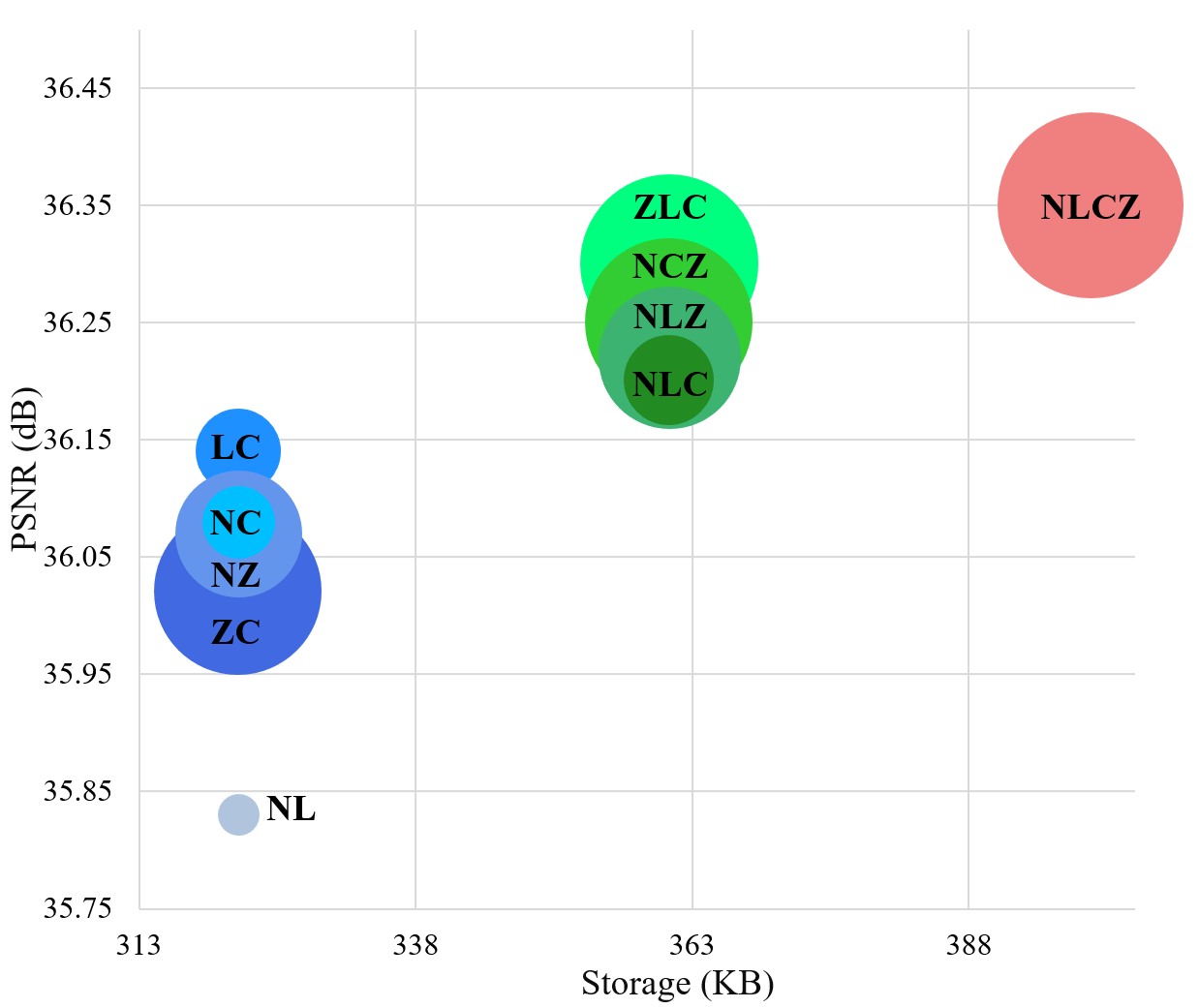}}
\end{minipage}
\caption{Performance-efficiency trade-off of different interpolation combinations with IM-LUT on $\times$2 SR of Set5. The diameter of each circle is proportional to the MACs for producing a 1280$\times$720 image. }
\label{fig_abl_visualization}
\end{figure}

To further analyze IM-LUT, \cref{fig_abl_visualization} illustrates the trade-off between performance and efficiency for different interpolation combinations on $\times2$ SR using Set5. More complex combinations, such as NLCZ, achieve higher PSNR at the expense of increased computational cost. In contrast, NLC and NLZ provide a more balanced trade-off between quality and efficiency. These results highlight IM-LUT’s capability to enhance SR performance while maintaining computational feasibility for resource-constrained applications.

\subsection{Visualization Results}
\begin{figure*}[!t] 
\centering
\begin{minipage}{1.0\linewidth}
\centerline{\includegraphics[width=0.98\textwidth, height=0.50\textheight]{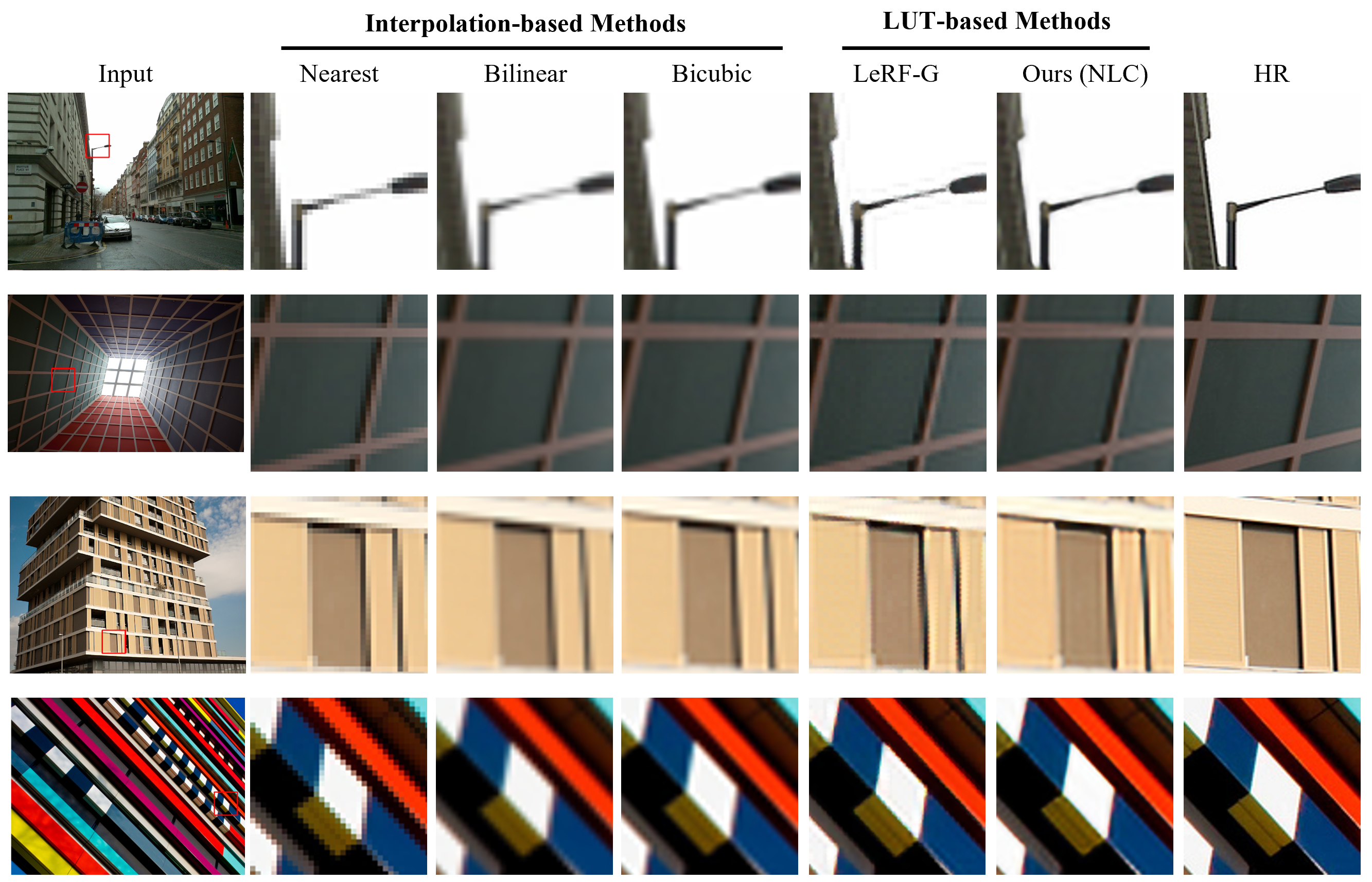}}
\end{minipage}
\caption{Comparison of $\times 3$ SR results of interpolation-based methods and LUT-based methods.} 

\label{fig_supple_srx3}
\end{figure*}

\begin{figure*}[!t] 
\centering
\begin{minipage}{1.0\linewidth}
\centerline{\includegraphics[width=0.95\textwidth, height=0.61\textheight]{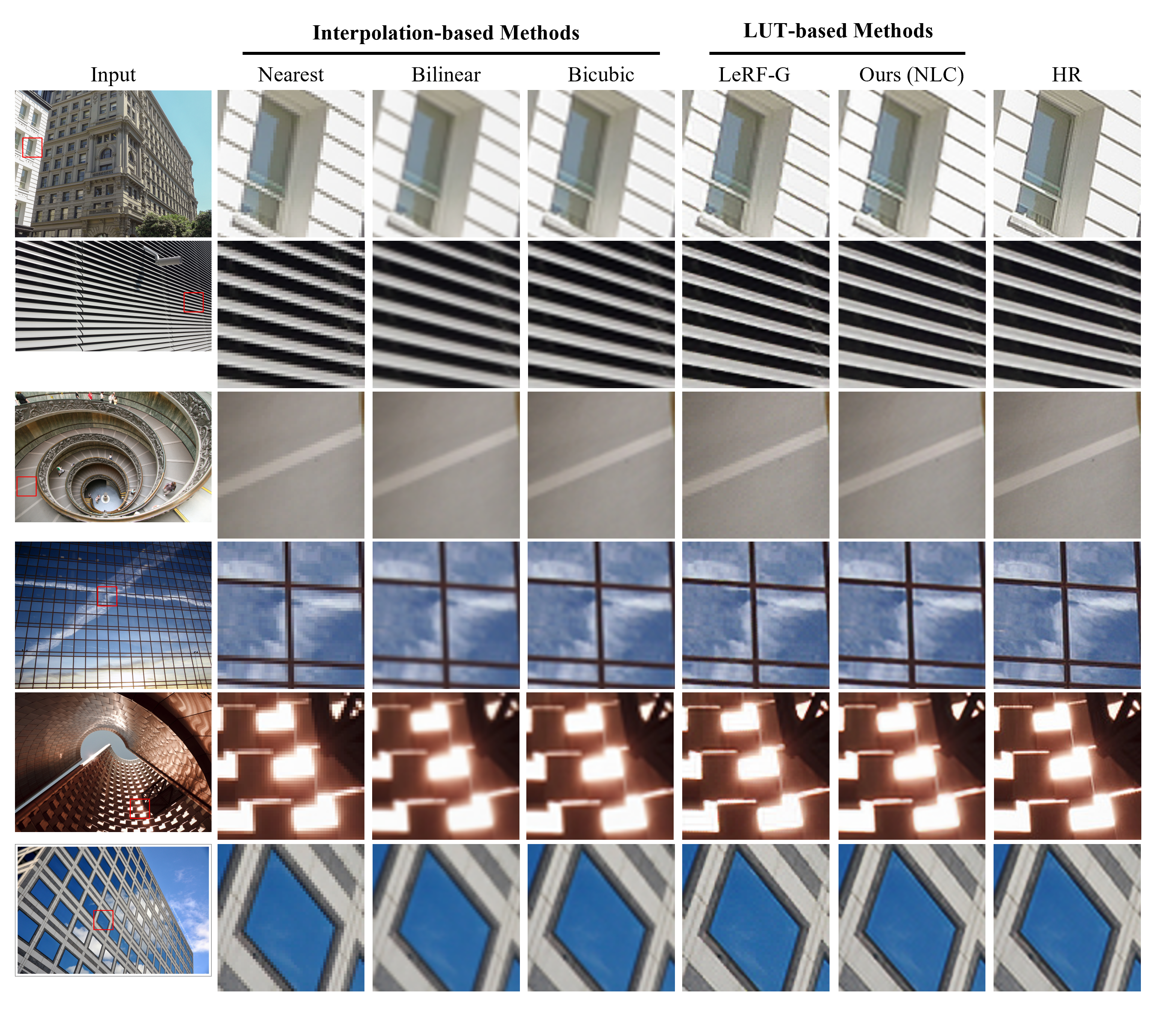}}
\end{minipage}
\vspace{-0.2cm}
\caption{Comparison of $\times 2$ SR results of interpolation-based methods and LUT-based methods.} 

\label{fig_supple_sr}
\end{figure*}

We further validate the effectiveness of IM-LUT through qualitative comparisons with interpolation-based methods and LeRF~\cite{li2023learning}. As shown in \cref{fig_supple_sr}, our approach achieves superior texture reconstruction and edge sharpness for $\times2$ SR on benchmark datasets compared to conventional interpolation and LUT-based SR models.

\begin{figure*}[!t] 
\centering
\begin{minipage}{1.0\linewidth}
\centerline{\includegraphics[width=0.95\textwidth, height=0.31\textheight]{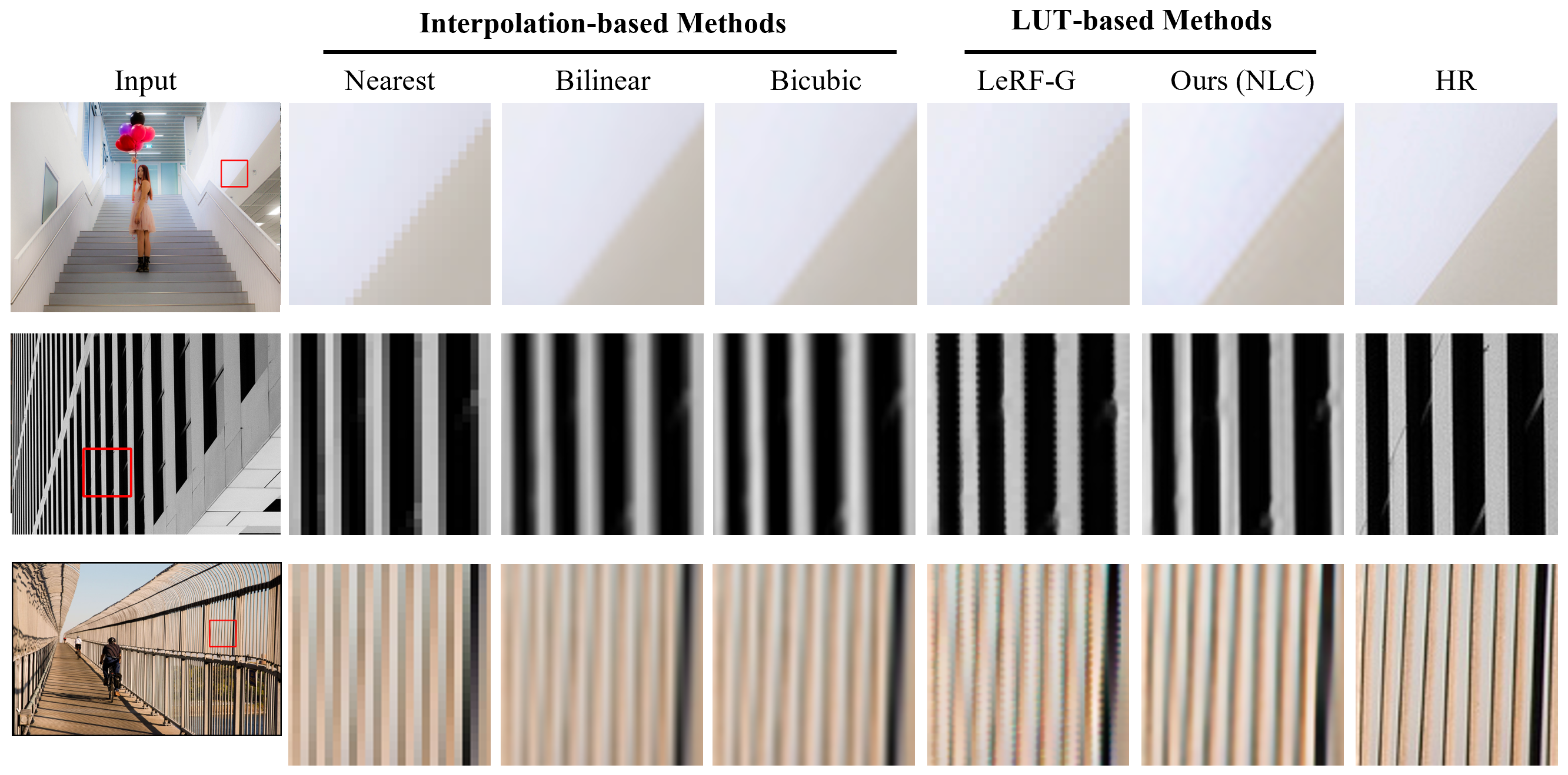}}
\end{minipage}
\vspace{-0.2cm}
\caption{Comparison of $\times 4$ SR results of interpolation-based methods and LUT-based methods.} 

\label{fig_supple_srx4}
\end{figure*}

Additionally, we extend our analysis to ASISR methods, as shown in \cref{fig_supple_srx3} and \cref{fig_supple_srx4}. IM-LUT consistently preserves text clarity and structural details at $\times3$ and $\times 4$ magnifications, effectively mitigating blurring and artifacts observed in LeRF-G~\cite{li2023learning}. This highlights the strength of our adaptive interpolation mixing strategy, ensuring stable and visually coherent results even beyond the training distribution.

\subsection{Explanation for Experiment Results}
In~\cref{fig_tradeoff}, the results confirm that IM-LUT achieves strong performance across arbitrary scales with a single LUT set. Compared to fixed-scale methods like MuLUT and SPF-LUT, which require separate models for each scale, IM-LUT generalizes to both seen and unseen (including non-integer) scaling factors without retraining. Even when evaluated at the fixed scales those baselines are optimized for, IM-LUT shows comparable accuracy while being significantly more storage-efficient. This supports the effectiveness of our interpolation-based design and its favorable cost-accuracy trade-off compared to existing LUT-based approaches.

\section{Details of IM-LUT}
\subsection{Rotation Ensemble}

In practice, to expand the receptive field, we apply a rotation ensemble strategy~\cite{jo2021practical} to both the weight predictor \( f_w(\cdot) \) and the refiner \( f_r(\cdot) \). The input patches are rotated by angles selected from the predefined set:

\begin{equation}
j \in \{ 0^\circ, 90^\circ, 180^\circ, 270^\circ \}.
\end{equation}

For each rotated input, \( f_w(\cdot) \) and \( f_r(\cdot) \) generate corresponding outputs, which are aggregated to obtain the final predictions:

\begin{equation}
\mathcal{W} = \frac{1}{J} \sum_{j} f_w ( R_j(\mathcal{I}_{{LR}}) ), \hspace{1em}
\bm{I}_{SR} = \frac{1}{J} \sum_{j}  f_r ( R_j(\bm{I}'_{{SR}}) ),  
\end{equation}
where $J$ denotes the number of selected rotations.

By leveraging multiple rotated versions of the input, this strategy effectively expands the receptive field without increasing storage requirements. 

Moreover, we employ a mixed-scale dataset for training, utilizing a total of seven scaling settings ranging from $\times 1.5$ to $\times 4.5$ in increments of $0.5$. This diverse scale range not only ensures that our model learns to handle arbitrary-scale super-resolution effectively but also enhances its ability to perform interpolation mixing by adapting to a wide variety of magnification factors. As a result, our method achieves improved flexibility and robustness in scale-adaptive SR.

\subsection{Refiner Architecture}
To enhance final output quality without increasing spatial resolution, we introduce a lightweight refiner designed with a directional ensemble structure. Unlike cascaded LUT architectures, our refiner operates in parallel, improving detail restoration while maintaining efficiency.

As the upscaling factor increases (e.g., $\times2 \rightarrow \times3 \rightarrow \times4$), the Refiner accounts for only a modest portion of the total computational cost (17\% to 20\% MACs), demonstrating its practical overhead. This design choice allows IM-LUT to remain competitive in both accuracy and efficiency, as further evidenced in our main results and ablation studies.

\end{document}